\documentclass[11pt]{article}

\usepackage{epsfig}
\usepackage{epstopdf}  
\usepackage{amsmath}
\usepackage[normalem]{ulem}

\renewcommand\sout{\bgroup \color{blue} \ULdepth=-.5ex \ULset}
\usepackage{color}
\usepackage{latexsym}
\usepackage{amssymb}
\usepackage{dsfont}
\usepackage{multirow}
\usepackage{epsfig}

\newcommand{\bea}{\begin{eqnarray}}
\newcommand{\eea}{\end{eqnarray}}
\newcommand{\be}{\begin{equation}}
\newcommand{\ee}{\end{equation}}

\newlength\savedwidth

\title{\bf{Pion transverse momentum dependent
parton distributions in the Nambu and Jona-Lasinio model}}

\author{Santiago Noguera\footnote{santiago.noguera@uv.es} \\
Departament de Fisica Te\`orica and IFIC, Universitat de Val\`encia
- CSIC,\\
E-46100 Burjassot, Spain 
\vspace{.3cm}
\\
Sergio Scopetta\footnote{sergio.scopetta@pg.infn.it}  \\
Dipartimento di Fisica e Geologia,
Universit\`a degli Studi di Perugia, 
\\ 
and Istituto Nazionale di Fisica Nucleare,
Sezione di Perugia, 
\\
via A. Pascoli, I - 06123 Perugia, Italy}	

\date{}

\begin{document}
\maketitle{}

\abstract{
An explicit evaluation of the two pion transverse momentum dependent 
parton distributions at leading twist is presented,
in the framework of the Nambu-Jona Lasinio model with Pauli-Villars
regularization.
The transverse momentum dependence of the obtained distributions is 
generated solely by the dynamics of the model. 
Using these results, the so called generalized Boer-Mulders shift
is studied and compared with recent lattice data.
The obtained agreement is very encouraging,
in particular because no additional parameter has been introduced.
A more conclusive comparison
would require a precise knowledge of the QCD evolution
of the transverse momentum dependent parton distributions
under scrutiny.
\vskip .5cm
KEYWORDS:
Deep Inelastic Scattering,
Phenomenological Models,
Chiral Lagrangians
}   



\section{Introduction}

The three-dimensional (3D) hadronic structure in
momentum space can be accessed through 
the transverse momentum dependent parton distributions
(TMDs) \cite{Collins:1981uw}, measured mainly in semi-inclusive
deep inelastic scattering (SIDIS) or in Drell-Yan (DY) processes.
For the nucleon target a large amount of theoretical work
is being done, driven by recent and forthcoming impressive
experimental efforts (see, e.g, \cite{Bacchetta:2006tn,Barone:2010zz,
Angeles-Martinez:2015sea} and references therein).
In this paper we discuss pion TMDs, which are experimentally
probed through the DY process (see, e.g., \cite{Peng:2014hta}
for a recent report).

At leading-twist, the pion structure is described in terms of two TMDs,
the unpolarized one, $f_{1,\pi}(x,k^{2}_T)$, 
describing the number density of partons with
longitudinal momentum fraction $x$ and transverse momentum $k_T$,
and the Boer-Mulders TMD, also called Boer-Mulders function, 
$h_{1,\pi}^{\perp }(x,k^2_T)$ 
\cite{Boer:1997nt,Boer:1999mm}.
The latter is not a density, being generated by spin-orbit correlations of 
transversely polarized partons; it is chiral-odd and therefore not accessible
in DIS, and it is ``naively'' time-reversal odd, i.e.,
under time reversal the correlation flips the sign.

TMDs are non perturbative quantities and they have not been calculated
from first principles, although 
recently lattice data
have been produced for the pion.
In particular, the lattice calculation in Ref. \cite{Engelhardt:2015xja}, 
performed at the pion mass $m_\pi = 518$ MeV, 
is the update of preliminary results 
reported in \cite{Engelhardt:2013nba,Musch:2011er}.
Pion TMDs have been estimated also in models of the pion
structure, such as spectator models 
\cite{Lu:2004hu,Burkardt:2007xm,Gamberg:2009uk}, 
bag models \cite{Lu:2012hh}, covariant model of the pion with Pauli-Villars
regulators, in the unpolarized case \cite{Frederico:2009fk},
and in a light-front constituent
quark model \cite{Pasquini:2014ppa}.

In this paper we present the calculation of  
$f_{1,\pi}(x,{k}^{2}_T)$ and
$h_{1,\pi}^{\perp }(x,{k}^{2}_T)$ in the model of Nambu and
Jona-Lasinio (NJL) \cite{Klevansky:1992qe}.

The NJL model is the most realistic model for the pseudoscalar mesons based on
a local quantum field theory built with quarks. It respects the realization of
chiral symmetry and gives a good description of low energy properties.
Mesons are described as bound states, in a fully covariant
way, using the Bethe-Salpeter amplitude, in a
field theoretical framework.
In this way, the Lorentz covariance of the problem is preserved.
The NJL model is a non-renormalizable field theory and therefore a cut-off
procedure has to be implemented. Here, the Pauli-Villars regularization
scheme has been chosen, because it respects the gauge symmetry of the
problem. The NJL model, together with its regularization procedure, can be 
regarded as an effective theory of QCD. 

The NJL model has a long tradition of successful predictions of different
observables related to the parton structure of pseudoscalar mesons, such as 
the parton distribution \cite{Davidson:1994uv,Davidson:2001cc},
generalized parton distributions
\cite{Theussl:2002xp},
distribution
amplitudes \cite{RuizArriola:2002bp}, transition distribution
amplitudes \cite{Courtoy:2007vy,CourtoyThesis}, transition form factors
\cite{Noguera:2010fe,Noguera:2011fv,Noguera:2012aw}.
Here, for the first time, we apply the same scheme to the calculation of
the pion TMDs. This will permit to obtain a 
dynamical $k_T$ dependence, at variance with various other model
analyses where its analytical trend was assumed, and to compare it
with very recent lattice data \cite{Engelhardt:2015xja}.

The paper is structured as follows. 
In Section 2 we describe our approach obtaining the formal results. In
the third Section we discuss the numerical results and,
at the end, we perform the comparison with lattice data. 
Conclusions are eventually presented in the last section.

\section{TMDs in the NJL model}

For a spinless particle, such as the pion, only two leading twist
TMDs arise, in contrast to the eight found for spin-$\frac{1}{2} $
particles \cite{Barone:2010zz}. The TMD $f_{1,\pi} $ is simply the unpolarized
quark distribution, whereas the Boer-Mulders (BM) function
\cite{Boer:1997nt},
$h_{1,\pi}^{\perp }$, describes the distribution of transversely polarized
quarks in the pion. The BM function is odd under time
reversal (T-odd). A non-zero value for this function is originated
by the final
and initial state interactions, in the SIDIS and DY processes,
respectively, which break the symmetry of the events under time
reversal. 

The calculation of $f_{1,\pi}$ and $h_{1,\pi}^{\perp }$ in the NJL model 
will be described in the following two subsections, respectively.

\subsection{Unpolarized TMD}

The unpolarized quark TMD in the pion is defined
as follows 

\begin{eqnarray}
f_{1,\pi}^{u(d)}\left(  x,k_{T}^{2}\right)   &  = &
\frac{1}{2}\int\frac{d\xi^{-}%
\,d^{2}\xi_{T}}{\left(  2\pi\right)  ^{3}}\,e^{-i\left(  \xi^{-}\,k^{+}%
-\vec{\xi}_{T}\vec{k}_{T}\right)  }
\nonumber
\\
& \times &  
\left\langle p\right\vert \bar{\psi}\left(  \xi^{-},\vec{\xi}_{T}\right)
\,\mathcal{L}_{\vec{\xi}_{T}}^{\dagger}\left(  \infty,\xi^{-}\right)
\,\gamma^{+}\,\frac{1}{2}\left(  1 \pm c\tau_{3}\right)  
\,\mathcal{L}_{0}\left(
\infty,0\right)  \,\psi\left(  0\right)  \left\vert p\right\rangle~,
\label{UqTMD.01}%
\end{eqnarray}

\noindent where $c=1(-1)$ stands for the $u(d)$ case, 
$k^{+}=xP^{+}$, $\xi^{+}=0,$ and
the gauge link is given by%

\begin{equation}
\mathcal{L}_{\vec{\xi}_{T}}\left(  \infty,\xi^{-}\right)  =\mathcal{P}%
\,\exp\left(  -ig_{s}\int_{\xi^{-}}^{\infty}A^{+}\left(  \eta^{-},\vec{\xi
}_{T}\right)  \,d\eta^{-}\right)~, 
\label{UqTMD.02}%
\end{equation}

\noindent with $g_{s}$ 
the strong $SU\left(  3\right)_{c}$ coupling constant. 

To fix the ideas, 
we consider the $u$ quark TMD in a $\pi^{+}.$ 
At zero
order in $g_{s}$, one gets%

\begin{equation}
f_{1,\pi}\left(  x,k_{T}^{2}\right)  =\frac{1}{2}%
\int\frac{d\xi^{-}\,d^{2}\xi_{T}}{\left(  2\pi\right)  ^{3}}\,e^{-i\left(
\xi^{-}\,k^{+}-\vec{\xi}_{T}\vec{k}_{T}\right)  }\,\left\langle p\right\vert
\bar{\psi}\left(  \xi^{-},\vec{\xi}_{T}\right)  \,\gamma^{+}\,\frac{1}%
{2}\left(  1 + \tau_{3}\right)  \,\psi\left(  0\right)  \left\vert
p\right\rangle~.
\label{uq2}%
\end{equation}

\noindent The two diagrams
contributing to this quantity are shown in Fig. 1. 
The contribution of the diagram in the left panel
is%

\begin{align}
f_{1,\pi}\left(  x,k_{T}^{2}\right)   &  =-\frac{1}%
{2}\int\frac{d^{4}q}{\left(  2\pi\right)  ^{4}}\delta\left(  k^{+}-\frac
{P^{+}}{2}-q^{+}\right)  \,\delta^{2}\left(  k_{T}-\frac{P_{T}}{2}%
-q_{T}\right)  \nonumber\\
&  \times\mathrm{Tr}\left(  iS_{F}\left(  q-\frac{P}{2}\right)  \,ig_{\pi
qq}\,\,i\gamma_{5} \, \tau_- \,iS_{F}\left(
q+\frac{P}{2}\right)  \right.  \nonumber\\
&  \times\left.  \gamma^{+}
\,iS_{F}\left(  q+\frac{P}{2}\right)  \,ig_{\pi qq}\,\,i\gamma_{5}\, \tau_+
\right)~,
\label{UqTMD.04}%
\end{align}

\noindent where \textrm{Tr} 
implies traces in color, flavor and Dirac matrices,
$S_F(p)$ is the Feynman propagator and $\tau_\pm = {1 \over \sqrt{2}}
(\tau_1 \pm i \tau_2)$. The
other diagram of Fig. 1, corresponding to the propagation of a $\sigma$
particle, which gives sometimes important contributions (see for instance
the calculation of pion GPDs in \cite{Theussl:2002xp}), 
vanishes in this case, where a diagonal
matrix element of a bi-local current is involved.

\begin{figure}
[ptb]
\begin{center}
\includegraphics[width=6.5cm]{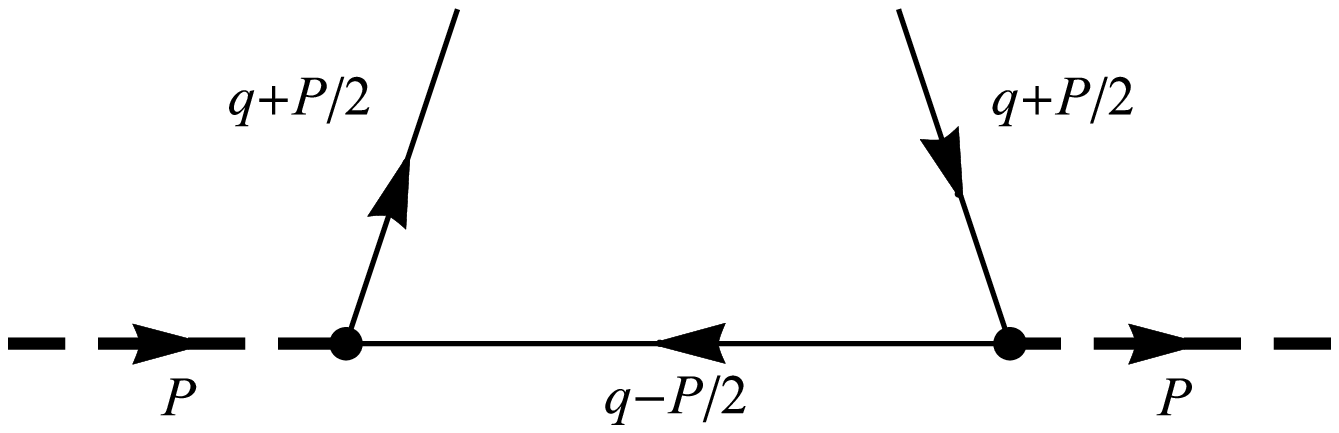}
\includegraphics[width=6.5cm]{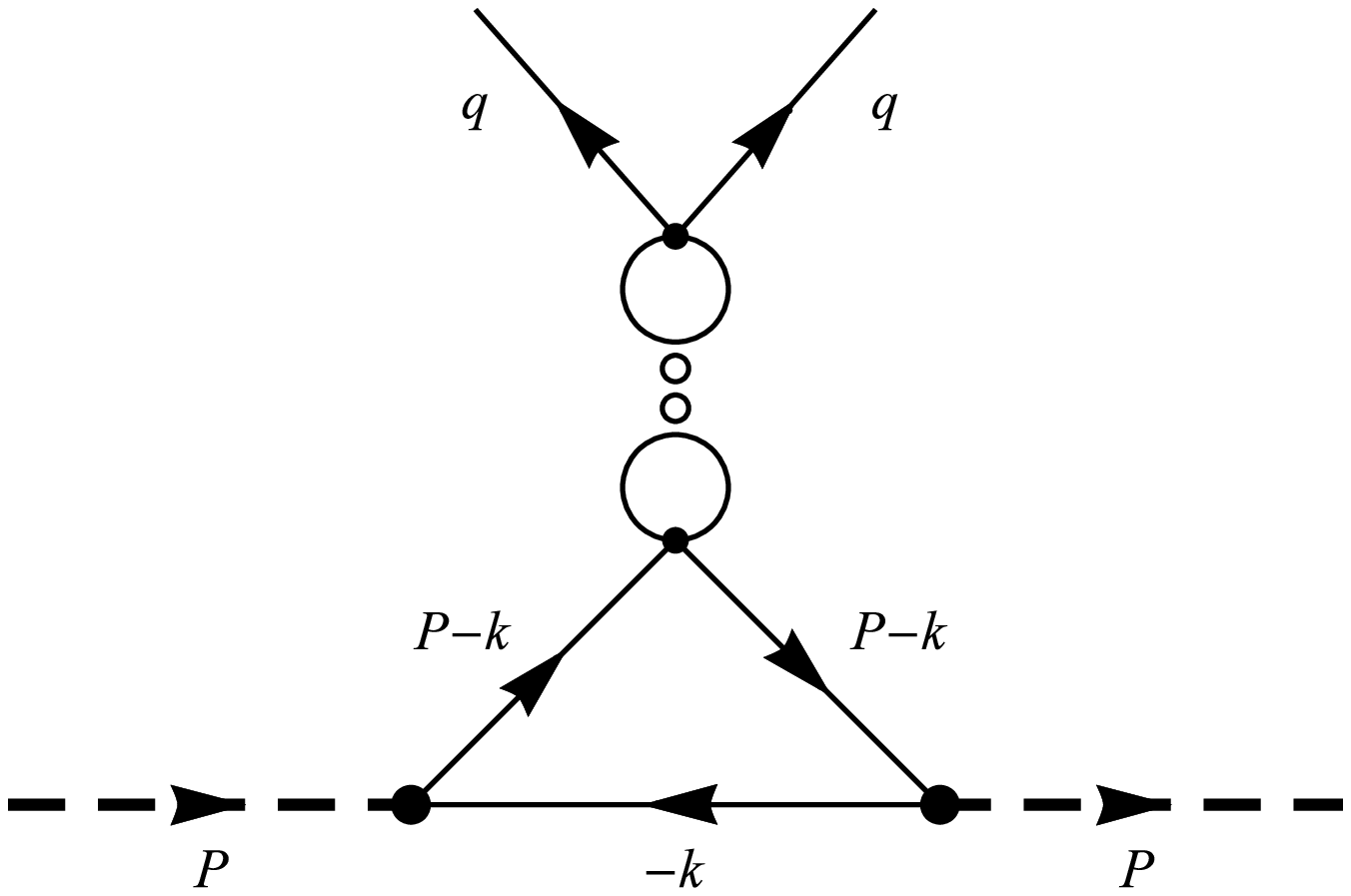}
\caption{The diagrams to be considered
in the evaluation of the $u$ TMD in a $\pi^+$, Eq. (\ref{uq2}).}%
\label{f1}%
\end{center}
\end{figure}

After Pauli Villars regularization (see the Appendix
for details), the final result is %
\begin{align}
f_{1,\pi}\left(  x,k_{T}^{2}\right)   &  =\frac
{3}{4\,\pi^{3}}\,g_{\pi qq}^{2}\,\theta\left(  x\right)  \,\theta\left(
1-x\right)  \,\sum_{i=0}^{2}c_{i}\nonumber\\
&  \times\,\left\{  \frac{1}{k_{T}^{2}+M_{i}^{2}-m_{\pi}^{2}\,x\left(
1-x\right)  }+\frac{m_{\pi}^{2}\,x\left(  1-x\right)  }{\left[  k_{T}%
^{2}+M_{i}^{2}-m_{\pi}^{2}\,x\left(  1-x\right)  \right]  ^{2}}\right\}  ~.
\label{final_unp}%
\end{align}

The integration over $k_{T}^{2}$ of the TMD
yields the pion PD

\begin{align}
q\left(  x\right)   &  =\int d^{2}k_{T}\,f_{1,\pi}\left(  x,k_{T}^{2}\right)
\nonumber\\
&  =\int\frac{d\xi^{-}}{2\pi}\,e^{-i\xi^{-}\,k^{+}}\left\langle p\right\vert
\bar{\psi}\left(  \xi^{-}\right)  \,\mathcal{L}_{0}^{\dagger}\left(
\infty,\xi^{-}\right)  \,\gamma^{+}\,\mathcal{L}_{0}\left(  \infty,0\right)
\,\psi\left(  0\right)  \left\vert p\right\rangle~,
\end{align}
with $k^{+}=xP^{+}$, $\xi^{+}=0$ and $\vec{\xi}_{T}=\vec{0}.$ 
One gets explicitly
\begin{eqnarray}
q\left(  x\right)  & = & \frac{3}{4\pi^{2}}\,g_{\pi qq}^{2}\,\theta\left(
x\right)  \,\theta\left(  1-x\right)  
\nonumber
\\
& \times &
\sum_{i=0}^{2}c_{i}\,\left\{  \ln
\frac{m^{2}-m_{\pi}^{2}\,x\left(  1-x\right)  }{\left(  M_{i}^{2}-m_{\pi}%
^{2}\,x\left(  1-x\right)  \right)  }+\frac{m_{\pi}^{2}\,x\left(  1-x\right)
}{\left[  M_{i}^{2}-m_{\pi}^{2}\,x\left(  1-x\right)  \right]  }\right\}~.
\label{qx}
\end{eqnarray}

We stress that, since we are working in a field theoretical scheme, the right
support of the distributions, $0\leq x\leq1$, is not imposed
and arises naturally. 
For the same reason, one can easily proof 
that: i) the normalization is correct, i.e., $\int dx~q\left(  x \right) = 1$;
ii) $\int dx~x~q\left(  x\right)  =0.5$, i.e.,
the fraction of momentum carried by each quark is one
half of the total momentum. Since at this level there are no sea
quarks, this is the expected correct result.

\subsection{Boer-Mulders function}

The BM function is defined as

\begin{eqnarray}
\label{BM.01}
h_{1,\pi}^{\perp \, u(d)}\left(  x,k_{T}^{2}\right)   
&  = &\epsilon^{i\,j\,}k_{T}%
^{j}\,\frac{m_{\pi}}{2\,k_{T}^{2}}\int\frac{d\xi^{-}\,d^{2}\xi_{T}}{\left(
2\pi\right)  ^{3}}\,e^{-i\left(  \xi^{-}\,k^{+}-\vec{\xi}_{T}\vec{k}%
_{T}\right)  }\\
& \times & \left\langle p\right\vert \bar{\psi}\left(  \xi^{-},\vec{\xi}_{T}\right)
\,\mathcal{L}_{\vec{\xi}_{T}}^{\dagger}\left(  \infty,\xi^{-}\right)
\,i\,\sigma^{i\,+}\,\gamma_{5}\,\frac{1}{2}\left(  1 \pm c\,\tau_{3}\right)
\,\mathcal{L}_{0}\left(  \infty,0\right)  \,\psi\left(  0\right)  \left\vert
p\right\rangle~,
\nonumber%
\end{eqnarray}
with the same conventions used in Eq. (\ref{UqTMD.01}).
To fix the ideas, 
as previously done for the unpolarized TMD, 
we will consider the BM function for a $u$ quark in a $\pi^{+}$.

At zero order in $g_{s}$, $h_{1,\pi}^{\perp}$ vanishes, due to the T-odd
character of the BM function.
In order to have a non-zero value of the BM function,
we expand the gauge link $\mathcal{L}_{\vec{\xi}_{T}}\left(  \infty,\xi
^{-}\right)  $, Eq. (\ref{UqTMD.02}) 
in powers of $g_{s}$, up to the first order, as
it has been done for phenomenological model estimates 
of T-odd parton distributions, for the nucleon (see, e.g., 
Refs. \cite{Brodsky:2002cx,Courtoy:2008vi,Courtoy:2008dn,Courtoy:2009pc,
Pasquini:2010af}) and, recently, for the pion
\cite{Pasquini:2014ppa}.
For
the gluonic field, we use its definition in terms of the source%
\begin{equation}
A^{a\,+}\left(  \eta\right)  =\int d^{4}y~D^{+\,\nu}\left(  \eta-y\right)
\,g_s\,\bar{\psi}\left(  y\right)  \,\frac{\lambda^{a}}{2}\,\gamma_{\nu}%
\,\psi\left(  y\right)~,
\label{gf}
\end{equation}
where $D^{\mu\,\nu}\left(  x-y\right)$
is the gluon propagator.
After some calculation, we arrive at
\begin{eqnarray}
\mathcal{L}_{\vec{\xi}_{T}}\left(  \infty,\xi^{-}\right)  
& \simeq &
1-g_{s}%
^{2}\,\frac{\lambda^{a}}{2}\int\frac{d^{4}t}{\left(  2\pi\right)  ^{4}%
}e^{i\,\vec{t}_{T}\vec{\xi}_{T}}\ e^{-i\,t^{+}\xi^{-}}~\frac{1}{t^{+}%
-i\varepsilon}\ 
\nonumber
\\
& \times &
\frac{-1}{t^{2}+i\varepsilon}\int d^{4}y~e^{i\,t\,y}%
\,\bar{\psi}\left(  y\right)  \,\frac{\lambda^{a}}{2}\,\gamma^{+}\,\psi\left(
y\right)~.
\end{eqnarray}

At the first order in $g_{s}^{2}$ one gets%
\begin{eqnarray}
h_{1,\pi}^{\perp}\left(  x,k_{T}^{2}\right)  
& = &
\epsilon^{i\,j\,}k_{T}%
^{j}\,\frac{m_{\pi}}{2\,k_{T}^{2}}\,g_{s}^{2}
\int\frac{d\xi^{-}\,d^{2}\xi_{T}}
{\left(  2\pi\right)^{3}}\,e^{-i\left(  \xi^{-}\,k^{+}-\vec{\xi}_{T}
\vec{k}_{T}\right)  }
\left\langle p\right\vert \bar{\psi}\left(  \xi^{-},\vec{\xi}_{T}\right)
\nonumber\\
& \times &
\,\left\{   \frac{\lambda^{a}}{2}\int\frac{d^{4}t}{\left(
2\pi\right)  ^{4}}e^{i\,\vec{t}_{T}\vec{\xi}_{T}}\ e^{-i\,t^{+}\xi^{-}}%
~\frac{1}{t^{+}-i\varepsilon}\ \frac{-1}{t^{2}+i\varepsilon} \right. 
\label{2body}
\\
& \times &
\left.
\int
d^{4}y~e^{i\,t\,y}\,\bar{\psi}\left(  y\right)  \,\frac{\lambda^{a}}%
{2}\,\gamma^{+}\,\psi\left(  y\right)  
\,i\,\sigma
^{i\,+}\,\gamma_{5}\,
\right.
\nonumber\\
& - &
i\,\sigma^{i\,+}\,\gamma_{5}\,
\left.  
\frac{\lambda^{a}}{2}\int\frac{d^{4}t}{\left(  2\pi\right)
^{4}}~\frac{1}{t^{+}-i\varepsilon}\ \frac{-1}{t^{2}+i\varepsilon}\int
d^{4}y~e^{i\,t\,y}\,\bar{\psi}\left(  y\right)  \,\frac{\lambda^{a}}%
{2}\,\gamma^{+}\,\psi\left(  y\right) \right\}  \,\psi\left(
0\right)  \left\vert p\right\rangle ~~.
\nonumber
\end{eqnarray}

\begin{figure}
[ptb]
\begin{center}
\includegraphics[width=6.5cm]{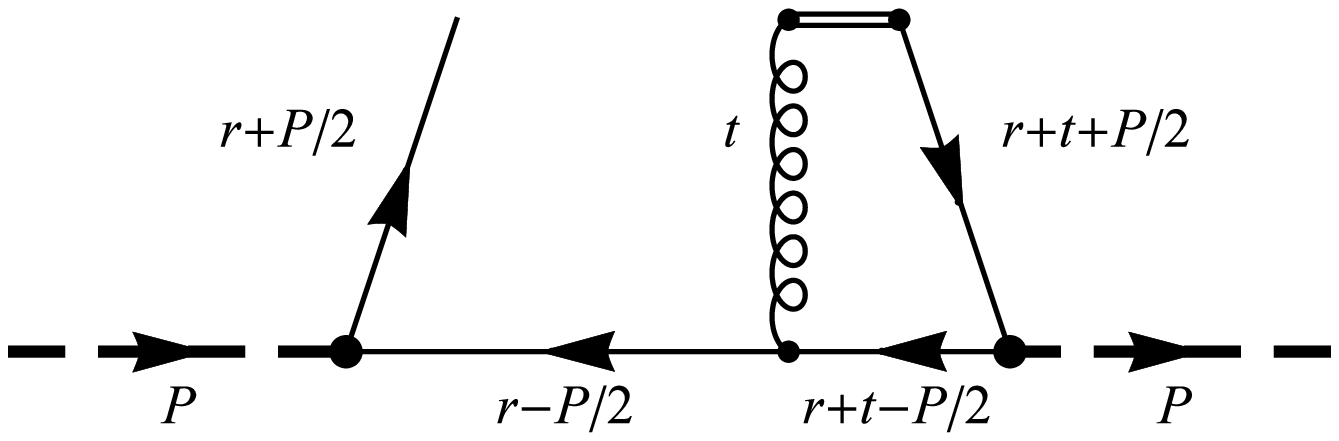}
\includegraphics[width=6.5cm]{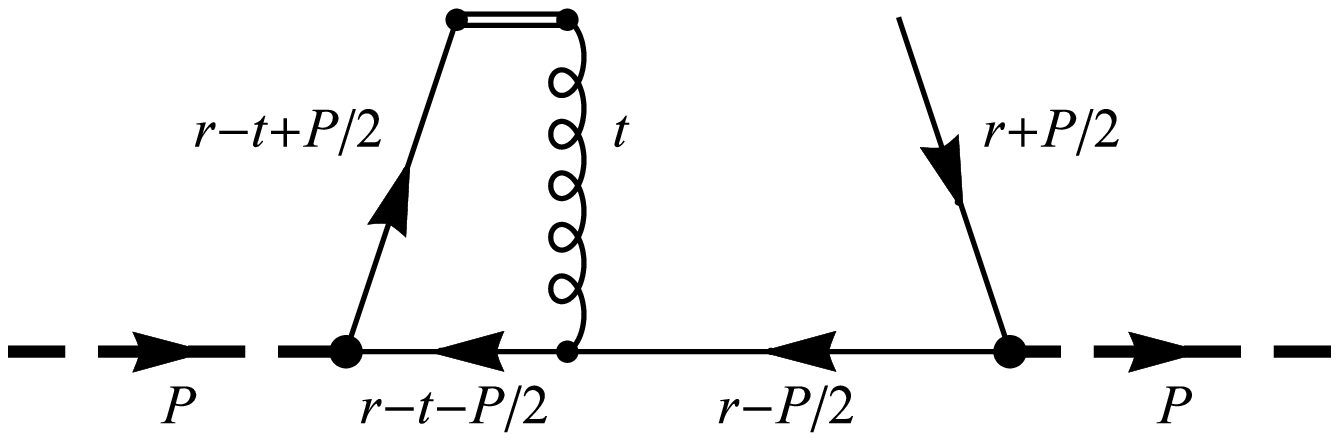}
\caption{
The diagrams describing the two traces in  Eq. (\ref{bm_tr}).
}%
\label{f7}%
\end{center}
\end{figure}

A straightforward calculation leads to
\begin{eqnarray}
h_{1,\pi}^{\perp}\left(  x,k_{T}^{2}\right)  & = &
-\epsilon
^{i\,j\,}k_{T}^{j}\,\frac{m_{\pi}}{2\,k_{T}^{2}}\,g_{s}^{2}\,g_{\pi qq}%
^{2}\,\int\frac{d^{4}t\,~dr^{-}}{\left(  2\pi\right)  ^{8}}\,\frac{1}%
{t^{+}-i\varepsilon}\,\frac{1}{t^{2}+i\varepsilon}
\nonumber
\\
& \times &
\left\{  -\mathrm{Tr}\left[  S_{F}\left(  {\small r+t-}\tfrac{P}%
{2}\right)  \gamma_{5}\, \tau_- \, S_{F}\left(
{\small r+t+}\tfrac{P}{2}\right)  \tfrac{\lambda^{a}}{2}i\sigma^{i\,+}%
\gamma_{5}
\right. \right. 
\nonumber
\\
& \times &
\left. \left.
S_{F}\left(  {\small r+}\tfrac{P}{2}\right) 
\gamma_{5}\, \tau_+ S_{F}\left(  {\small r-}\tfrac{P}{2}\right)  \tfrac{\lambda^{a}}{2}%
\gamma^{+}\right]  \right.  
\nonumber
\\
& + &
\left.  \mathrm{Tr}\left[  S_{F}\left(  {\small r-}\tfrac{P}{2}\right)
\gamma_{5} \, \tau_-  \, S_{F}\left(  {\small r+}%
\tfrac{P}{2}\right)  \tfrac{\lambda^{a}}{2}i\sigma^{i\,+}\gamma_{5}
\right. \right.
\nonumber
\\
& \times &
\left. \left.
S_{F}\left(  {\small r-t+}\tfrac{P}{2}\right)  \gamma_{5} \, \tau_+%
S_{F}\left(  {\small r-t-}\tfrac{P}{2}\right)  \tfrac{\lambda^{a}}{2}%
\gamma^{+}\right]  \right\}  ~,\label{bm_tr}%
\end{eqnarray}
where 
$r=\left(  \left(  x-\frac{1}{2}\right)  P^{+},\vec{k}_{T},r^{-}\right).$ 
The traces in the equation above, in the order they appear, correspond to
the diagrams in the left and right panels of
Fig. \ref{f7}, respectively. In
principle, the BM function could have contributions also from the
sigma term (the one reported in Fig. \ref{f1}, right panel, 
in the unpolarized case). The
direct calculation shows anyway that these contributions vanish. %

After a lengthy calculation, 
and including the Pauli Villars renormalization,
we obtain
\begin{eqnarray}
h_{1,\pi}^{\perp}\left(  x,k_{T}^{2}\right)  & = &
- \frac{1}{\sqrt{2}\,\pi^{3}%
}\,\frac{m\,m_{\pi}}{k_{T}^{2}}\,\alpha_s\,g_{\pi qq}^{2}\,\theta\left(
x\right)  \,\theta\left(  1-x\right)  \,\sum_{i=0}^{2}c_{i}\,\frac{1}%
{k_{T}^{2}+M_{i}^{2}-m_{\pi}^{2}\,x\left(  1-x\right)  }
\nonumber
\\
& \times &
\ln\left[  \frac
{k_{T}^{2}+M_{i}^{2}-m_{\pi}^{2}\,x\left(  1-x\right)  }{M_{i}^{2}-m_{\pi}%
^{2}\,x\left(  1-x\right)  }\right]~,
\label{bmf}
\end{eqnarray}
where we have introduced the strong coupling constant $\alpha_s = {g_s^2
\over 4 \pi }$.

The $k_{T}$ integral can be calculated, providing
\begin{eqnarray}
h_\pi (x) & = &
\int d^{2}k_{T}\,h_{1,\pi}^{\perp}\left(  x,k_{T}^{2}\right)  =-\frac
{1}{6\,\sqrt{2}}\,m\,m_{\pi}\,\alpha_{s}\,g_{\pi qq}^{2}%
\,\theta\left(  x\right)  \,\theta\left(  1-x\right)  
\nonumber
\\
& \times &
\,\sum_{i=0}^{2}%
c_{i}\,\frac{1}{M_{i}^{2}-m_{\pi}^{2}\,x\left(  1-x\right)  }~,%
\label{mome}
\end{eqnarray}
which, integrated over $x$, yields%
\begin{equation}
\int dx\int d^{2}k_{T}\,h_{1,\pi}^{\perp}\left(  x,k_{T}^{2}\right)
=-\frac{1}{6\,\sqrt{2}}\,m\,\alpha_{s}\,g_{\pi qq}^{2}\,\sum
_{i=0}^{2}c_{i}\,\frac{4\operatorname{arccsc}\left(  2\frac{M_{i}}{m_{\pi}%
}\right)  }{\sqrt{\left(  4\,M_{i}^{2}-m_{\pi}^{2}\right)  }}~.%
\end{equation}
In the $m_{\pi}\rightarrow0$ limit, we get%
\begin{equation}
\lim_{m_{\pi}\rightarrow0}\int dx\int d^{2}k_{T}\,h_{1,\pi}^{\perp}\left(
x,k_{T}^{2}\right)  =-\frac{1}{6\,\sqrt{2}}\,m_{\pi}\,m\,\alpha_{s}
\,g_{\pi qq}^{2}\,\sum_{i=0}^{2}\frac{c_{i}}{M_{i}^{2}}~.%
\label{inu}
\end{equation}

\noindent At variance with the $f_{1, \pi}(x, k_T^2)$ case, in which
$\int dx \, d k_T^2 \, f_{1 , \pi}(x, k_T^2) =1$ is a consequence of
charge conservation, the quantity (\ref{inu}) is not
in general related to any physical observable.
However we note that, in the present NJL framework and in the chiral limit,
one has $g_{\pi qq}^{2}\,\sum_{i=0}^{2}\frac{c_{i}}{M_{i}^{2}}=
{4 \over 3} \pi^2 r_\pi^2$, i.e.,
the right hand side of Eq. (\ref{inu}) can be related to $r_\pi$, the
charge radius of the pion.

\section{Discussion and comparison with lattice data}

\subsection{Unpolarized TMD}

\begin{figure}
[ptb]
\begin{center}
\includegraphics[width=7.cm]{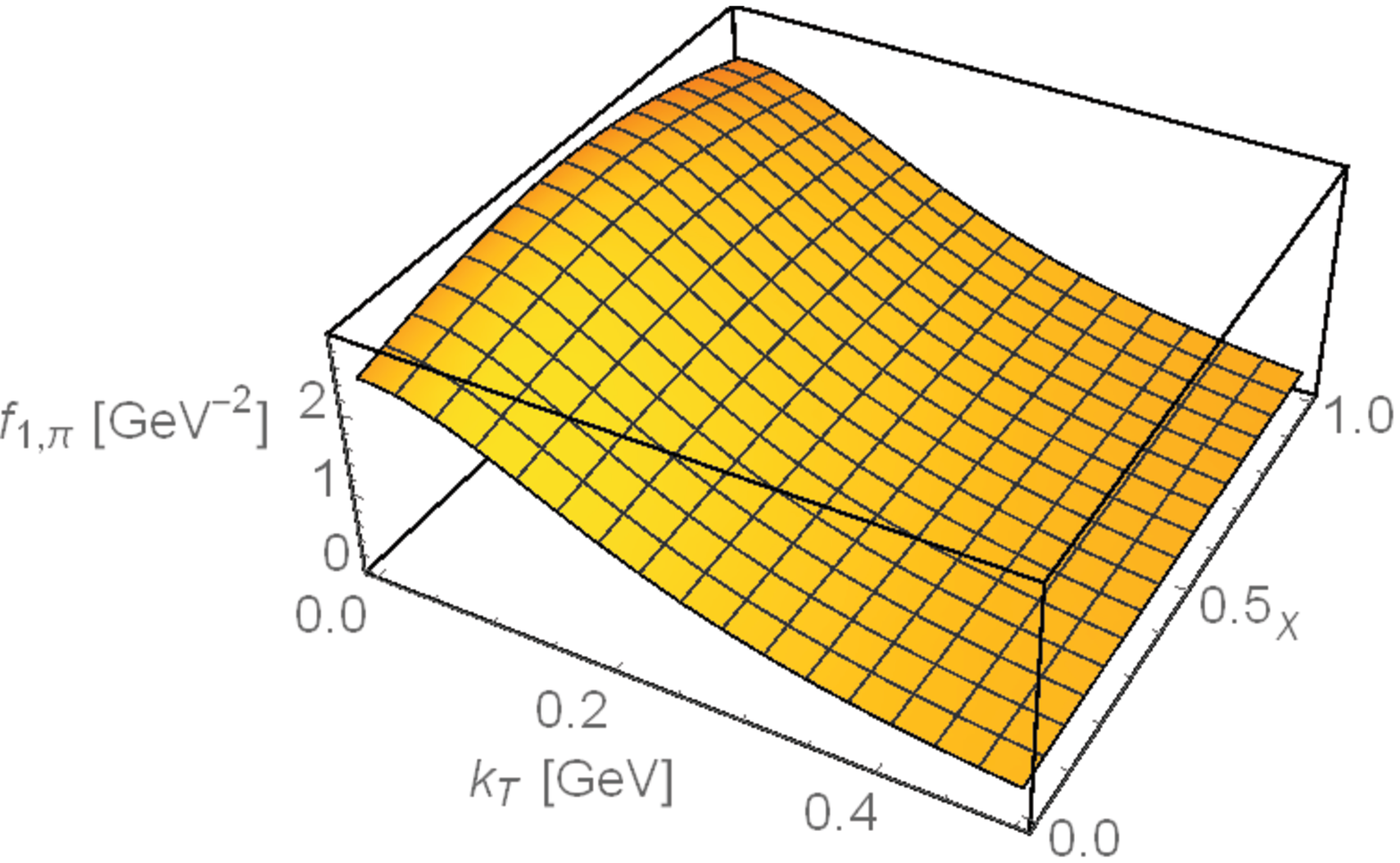}
\hspace{0.5cm}
\includegraphics[width=7.cm]{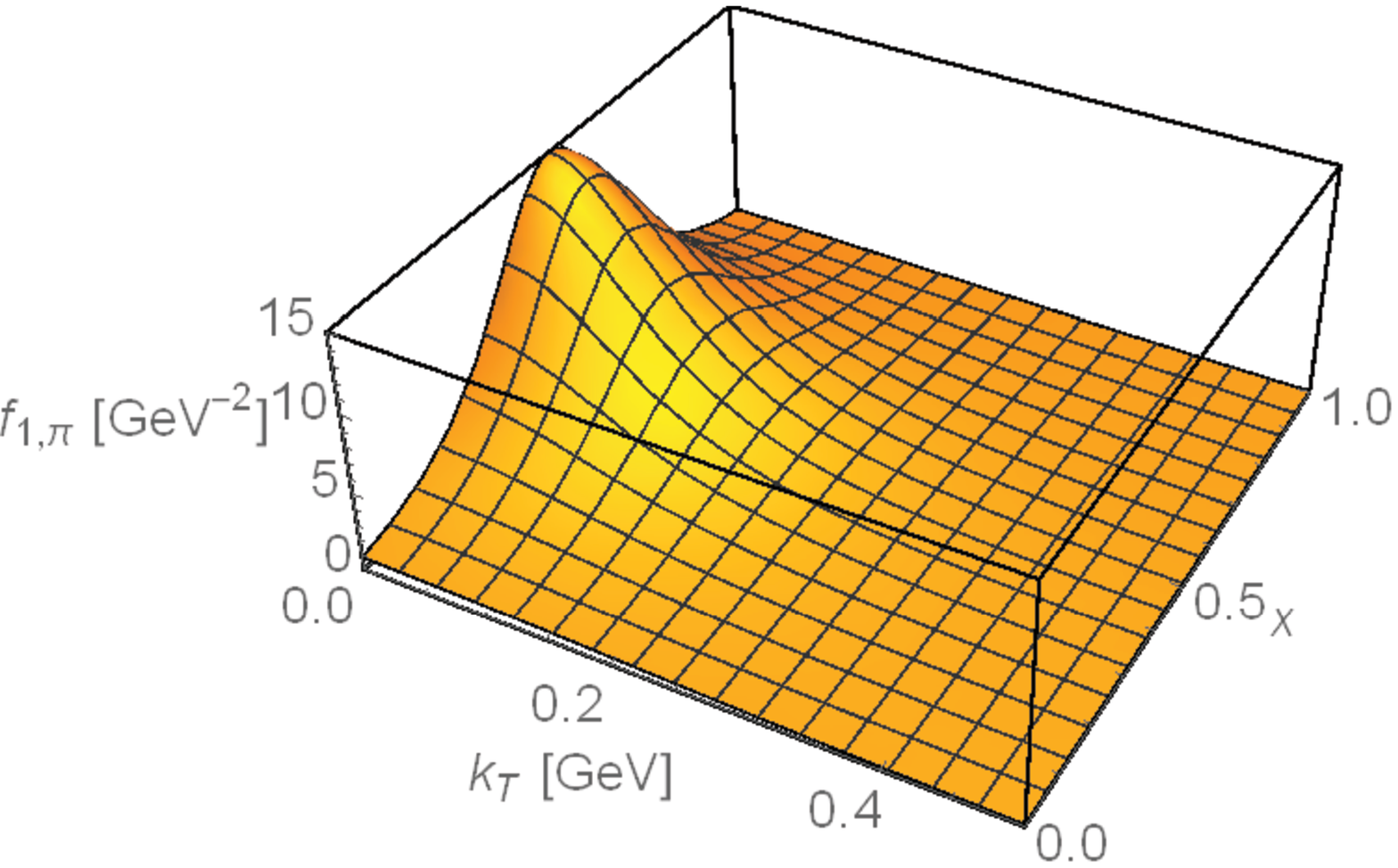}
\caption{
The unpolarized TMD
$f_{1,\pi}\left(x,k_{T}^{2}\right)$,
as a function of $k_{T}$ and $x$,
with $m_\pi=140$ MeV (left panel) and $m_\pi=518$ MeV (right panel).
}%
\label{f4_1}%
\end{center}
\end{figure}

To have a pictorial representation of the global $x$ and
$k_{T}$ dependencies, 3D-plots are shown in Fig.
\ref{f4_1}, for $m_\pi=$ 140 MeV (left panel) and $m_\pi=518$ MeV 
(right panel). In the left panel, 
it can be seen that the unpolarized TMD
varies slowly with $x$. This is easily understood looking at 
Eq. (\ref{final_unp}), where $x$ dependent terms
always appear multiplied by $m_\pi^2$.
In the right panel of Fig. \ref{f4_1} it is clearly seen that,
by taking a heavy pion with $m_\pi$ = 518 MeV, 
a value which will be useful later
for the comparison with lattice data, the $x$ dependence becomes
much more pronounced.
In the latter situation, our results agree qualitatively with the
findings of Ref. \cite{Frederico:2009fk,Pasquini:2014ppa}, where 
different constituent quark models have been used to
evaluate the unpolarized TMDs. This fact can be
understood thinking that, in the present NJL approach, the chiral limit
is naturally included, at variance with a constituent quark scenario,
where chiral symmetry is explicitly broken.
As a consequence, the $x$ dependence of our results 
with a pion mass of $m_\pi=$ 518 MeV
is closer to that obtained within constituent quark models,
with respect to what is obtained in our approach 
using the physical pion mass. 
\begin{figure}
[ptb]
\begin{center}
\includegraphics[width=9cm]{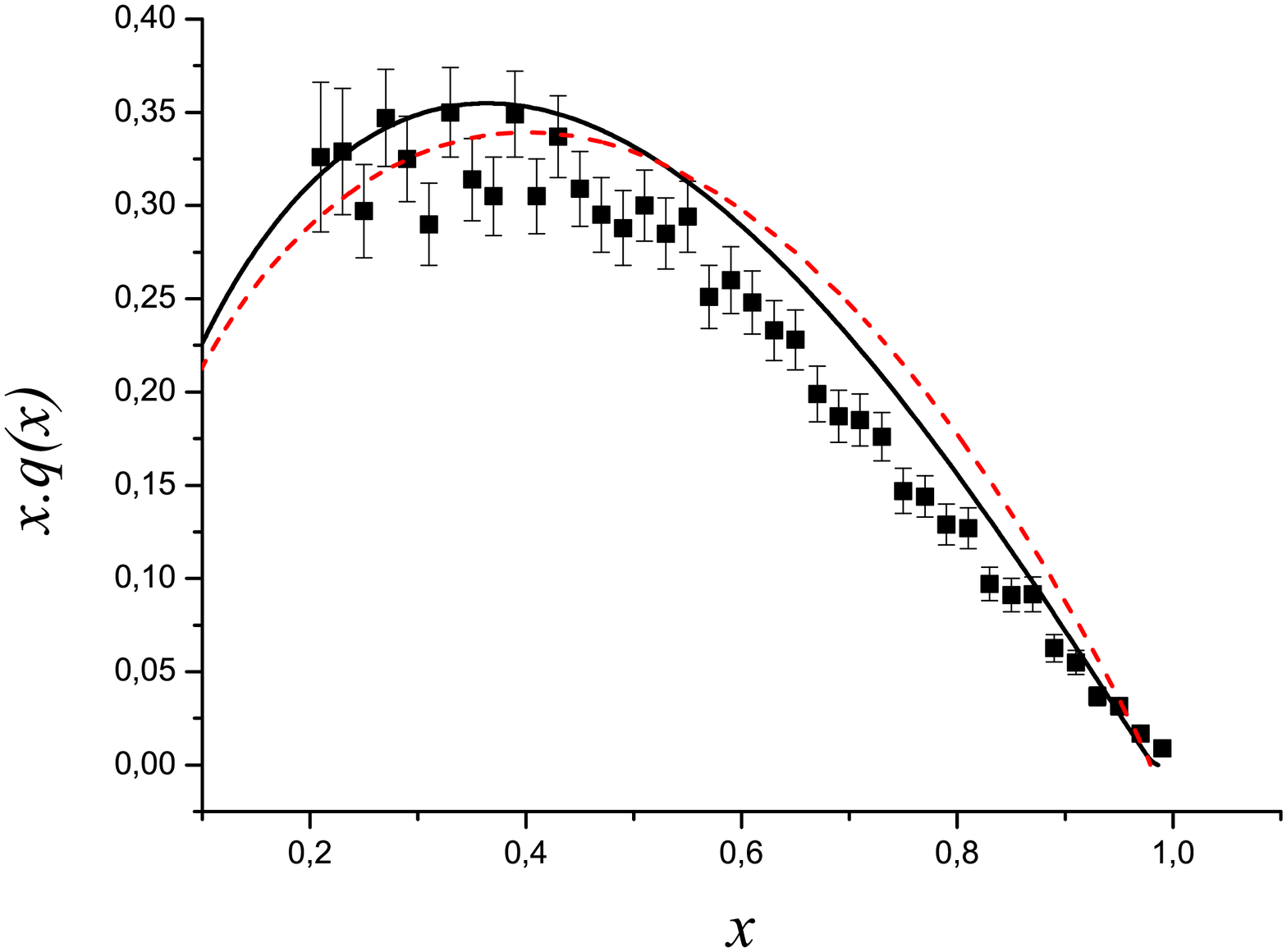}
\caption{
The pion PD, Eq. (\ref{qx}), after LO (solid line)
and NLO (dashed line) QCD evolution
to $Q=2$ GeV (picture taken from Ref. \cite{CourtoyThesis}).
Data are from Ref. \cite{Conway:1989fs}.
}%
\label{bonita}%
\end{center}
\end{figure}

The rather flat $x$ dependence, obtained
using $m_\pi = 140$ MeV, is not a drawback of the model. 
Hadron models, like the NJL model, must be regarded as a realization 
of QCD at a very low $Q^2$. Evolution will change
the $x$ dependence
in an important way. In fact, starting from Eq (\ref{qx}),
in Refs. \cite{Broniowski:2007si}
and \cite{CourtoyThesis,Courtoy:2008nf},
a very good description of the data
of the pion parton distribution at $Q=2$ GeV 
\cite{Conway:1989fs}
is obtained, as one can see in Fig. \ref{bonita},
taken from  \cite{CourtoyThesis}.
For later convenience, it is useful to report
that the LO parameters of the QCD evolution used
in Refs. \cite{Broniowski:2007si}
and \cite{CourtoyThesis,Courtoy:2008nf}
predict
$\alpha_s($2 GeV$)$ = 0.32 and $\alpha_s($2 GeV$)$ = 0.29,
respectively.
These values are in good agreement with 
$\alpha_s$ measured 
in correspondence of the mass of the $\tau$ lepton,
$\alpha_s(m_\tau=1.777$ GeV$) =0.327^{+0.019}_{-0.016}$
\cite{Agashe:2014kda}.

Concerning the relation between mass and $x$ dependence,
it is also interesting to observe that, in the chiral limit, 
the NJL model predicts an absolutely flat parton distribution, 
$q_\chi (x)=1$, and distribution amplitude, $\phi_\chi (x)=1$. 
Nevertheless, the different regime of evolution (DGLAP for the first 
quantity and ERBL for the second one) produces very different 
$x$ dependencies at higher $Q^2$ \cite{Noguera:2010fe}.

\begin{figure}
[ptb]
\begin{center}
\includegraphics[width=9cm]{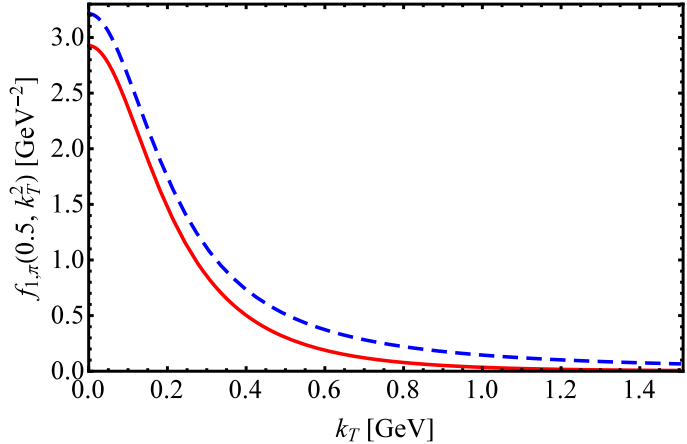}
\caption{
The unpolarized TMD
$f_{1,\pi}\left(x,k_{T}^{2}\right)$,
Eq. (\ref{final_unp}),  
at $x=0.5$, for the physical pion,
as a function of $k_{T}$ (full). The dashed curve does not include
the counter terms coming from the regularization procedure.
}%
\label{f4}%
\end{center}
\end{figure}

In Fig. \ref{f4}, the  $k_{T}$ dependence
is shown, having fixed $x=0.5$.
The result without 
the contributions of the counter terms originated by the 
regularization procedure is also reported.
It is worth stressing
that, in our approach, the  $k_{T}$ dependence is 
automatically generated by the NJL
dynamics.
This is an important feature of our results,
not found in other approaches.
In facts, for example, the two different $k_T$ dependencies 
of the unpolarized TMD
shown in Ref. \cite{Frederico:2009fk} are dictated by two different
forms adopted for a regulator function appearing in the pion Bethe-Salpeter
amplitude. In a similar fashion, in the Light-Front scenario of
Ref. \cite{Pasquini:2014ppa}, the
obtained $k_T$-dependence is determined by the gaussian
form assumed in the pion light-cone wave function,
following Refs. \cite{Schlumpf:1994bc,Chung:1988mu}.
In our case, the $k_T$ dependence is not imposed using an educated
guess.
It is therefore relevant to report
that our prediction has the following
asymptotic $k_T$ behavior, as it can be obtained
from Eq. (\ref{final_unp}):
\begin{eqnarray}
f_{1,\pi}\left(x,k_{T}^{2}\right)
\xrightarrow[k_T^2 \to \infty]{}  
\frac{3 g_{\pi qq}^2 \Lambda^4}{2 \pi^3 k_T^6}~.
\end{eqnarray}

\noindent
We reiterate that this is just a consequence of the NJL model
with Pauli-Villars regularization. 
To this respect, Fig. \ref{f4} points out the importance of the 
regularization procedure.
In facts, without the counter terms, which suppress the high
$k_{T}$ region, the TMD would not be integrable in the variable
$k_{T}^{2}$. 
Actually, as it has been shown above,
the integration over $k_{T}^{2}$ of the TMD 
yields the pion PD.

\subsection{Boer-Mulders function}

\begin{figure}
[ptb]
\begin{center}
\includegraphics[width=7.5cm]{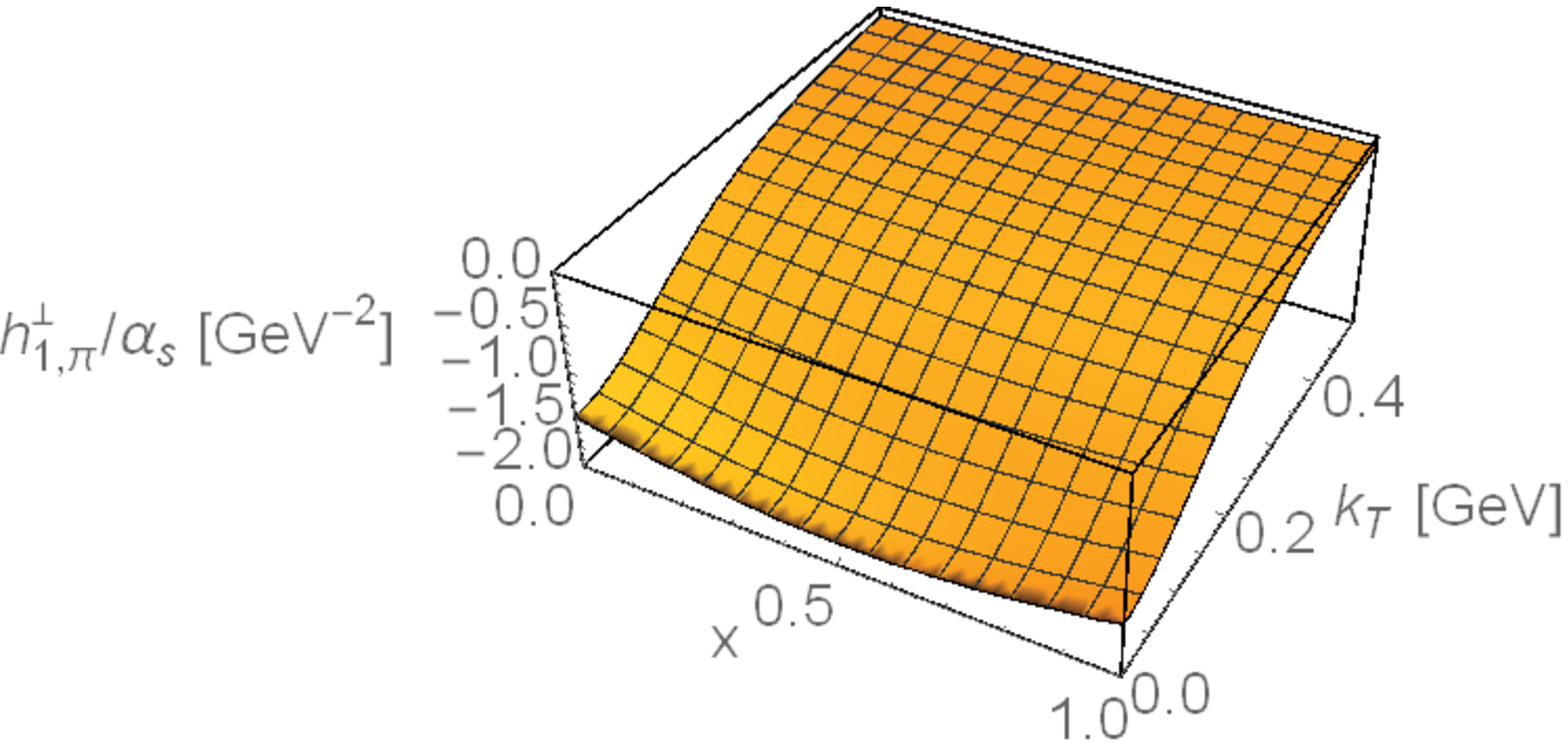}
\includegraphics[width=7.5cm]{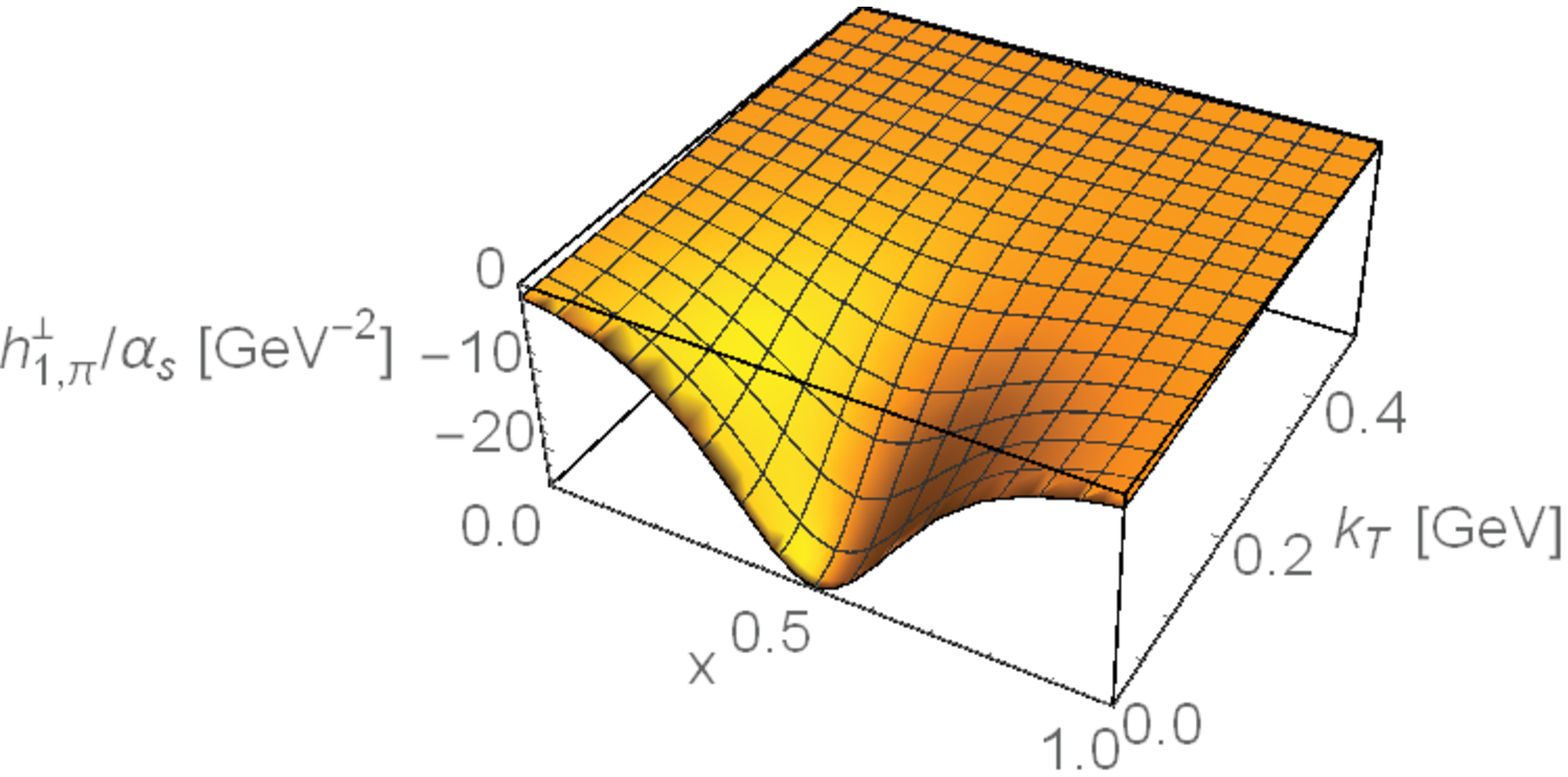}
\caption{
The Boer-Mulders TMD,
$h_{1,\pi}^{\perp}\left(x,k_{T}^{2}\right)$,
divided by the strong coupling constant
$\alpha_{s}$,
evaluated
using $m_\pi=140$ MeV (left panel) and $m_\pi=518$ MeV (right panel).
}%
\label{f11_1}%
\end{center}
\end{figure}

Numerical results of the evaluation of the BM function,
Eq. (\ref{bmf}), divided by the strong coupling constant,
are reported in Fig. \ref{f11_1}, in a 3D-plot,
providing a pictorial representation of the global $x$ and
$k_{T}$ dependencies.

As it happens for the unpolarized TMD, the BM TMD
varies slowly with $x$ when the physical pion
mass, $m_\pi$ = 140 MeV, is used in our calculation.
This is again easily understood looking at 
Eq. (\ref{bmf}), where $x$ dependent parts
always appear multiplied by $m_\pi^2$.
In the right panel of Fig. \ref{f11_1}, it is shown that,
by taking $m_\pi$ = 518 MeV, the $x$ dependence becomes
much more relevant.

As for the unpolarized TMD, the obtained $k_T$ behavior
is a genuine result of the NJL dynamics with
Pauli-Villars regularization. 
We report
our prediction for the 
asymptotic $k_T$ behavior of the BM TMD, as it can be obtained
from Eq. (\ref{bmf}):
\begin{eqnarray}
h_{1,\pi}^{\perp}\left(x,k_{T}^{2}\right)
& \xrightarrow[k_T^2 \to \infty]{} &  
\theta(x) \theta(1-x) \,
\frac{m m_\pi \alpha_s g_{\pi qq}^2}{ \sqrt{2} \pi^3 k_T^4}
\, \sum_{i=0}^2 \, c_i \, log{ M_i^2 - m_\pi^2 x (1-x) \over
m^2 - m_\pi^2 x (1-x) } 
\nonumber
\\
& + & O \left ( { 1 \over k_T^6} log ( k_T ) \right )
\end{eqnarray}

\subsection{Comparison with lattice data}

In the following, we compare our results with 
lattice measurements.
In facts, very recently, a lattice calculation
has been performed \cite{Engelhardt:2015xja},
focused on
a TMD observable related to the Boer-Mulders
effect in a pion.

The quantity which has been addressed is
a ratio, defined in an appropriate way to be safely
evaluated on the lattice.
It is the so called ``generalized Boer-Mulders shift'', given
by the following expression
\begin{equation}
\langle k_y \rangle_{UT} (b_T^2 ) \equiv m_{\pi }
\frac{\tilde{h}_{1}^{\perp [1](1)}
(b_T^2 )}{\tilde{f}_{1}^{[1](0)} (b_T^2 )}~,
\label{gbmshift}
\end{equation}

\noindent where ${\tilde{h}_{1}^{\perp [1](1)}
(b_T^2 )}$ and ${\tilde{f}_{1}^{[1](0)} (b_T^2 )}$
are $x$-moments of generic 
Fourier-transformed TMDs:

\begin{eqnarray}
\tilde{f}^{[m](n)} (b_T^2 ) &=& n! \left( -\frac{2}{m_\pi^2 }
\partial_{b_T^2 } \right)^{n} \int_{-1}^{1} dx\, x^{m-1}
\int d^2 k_T \,
e^{ib_T \cdot k_T } \ f(x,k_T^2 )
\label{eq10}
\\
&=& 
\frac{2\pi n!}{(m_\pi^2)^n}\, \int_{-1}^{1} dx\, x^{m-1}
\int d |k_T | \, |k_T | \,
\left( \frac{|k_T |}{|b_T |} \right)^n 
J_n(|b_T | |k_T |) f(x, k_T^2  )~,
\nonumber
\end{eqnarray}
with $J_n $ denoting the Bessel functions of the first kind.

One should notice that the $b_T \rightarrow 0$ limit of
the generalized Boer-Mulders shift, Eq. (\ref{gbmshift}), 
formally corresponds to $k_T^{2} $-moments of TMDs,
\begin{eqnarray}
\tilde{f}^{[m](n)} (0) &=& \int_{-1}^{1} dx\, x^{m-1}
\int d^2 k_T \ \left( \frac{k_T^2 }{2m_\pi^2 }\right )^{n} \
f(x,k_T^2 )
\label{ktmom}
\end{eqnarray}
\begin{figure}
[ptb]
\begin{center}
\includegraphics[width=9cm]{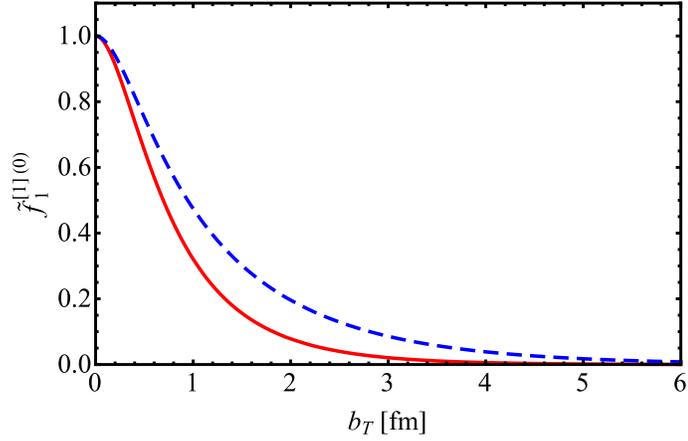}
\caption{
The moment ${\tilde{f}_{1}^{[1](0)} (b_T^2 )}$ 
of the unpolarized pion TMD,
calculated assuming the physical pion mass,
$m_\pi=140$ MeV (full line), or $m_\pi=518$ MeV, the value
used in the lattice calculation in Ref. \cite{Engelhardt:2015xja} 
(dashed line).
}%
\label{f13}%
\end{center}
\end{figure}
\begin{figure}
[ptb]
\begin{center}
\includegraphics[width=9cm]{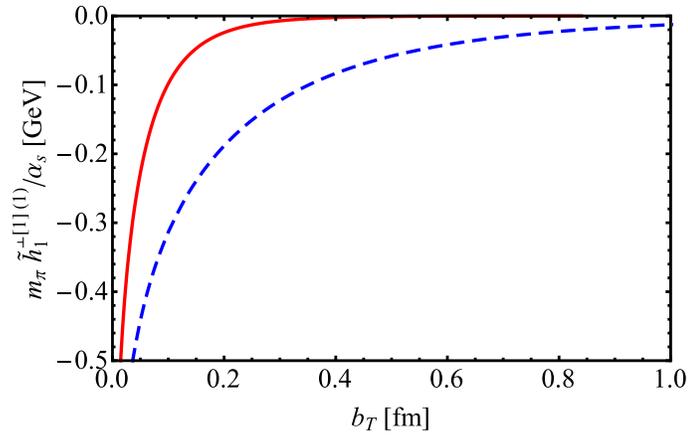}
\caption{
The same as in Fig. \ref{f13}, for $m_\pi$ times the moment
$\tilde{h}_{1}^{\perp [1](1)}(b_T^2 )$
of the Boer-Mulders pion TMD, divided by $\alpha_s$.
}%
\label{f14}%
\end{center}
\end{figure}
In the $b_T \rightarrow 0$ limit, the generalized
Boer-Mulders shift reduces therefore to the ``Boer-Mulders shift'',
\begin{equation}
\langle k_y \rangle_{UT} (0) = 
m_{\pi } \frac{\tilde{h}_{1}^{\perp [1](1)}
(0)}{\tilde{f}_{1}^{[1](0)} (0)}~,
\label{bmshift}
\end{equation}
which has the meaning of
the average transverse momentum in $y$-direction of quarks
polarized in the transverse (``$T$'') $x$-direction, in an unpolarized
(``$U$'') pion, normalized to the corresponding number of valence quarks.

It should be noted, however, that the
$k_T^{2} $-moments of TMDs (\ref{ktmom}) appearing in (\ref{bmshift})
are in general divergent 
at large $k_T $ and thus
not well-defined without an additional regularization. 
In the generalized quantity,
(\ref{gbmshift}), a finite $b_T $
effectively acts as a regulator through the associated Bessel weighting,
cf.~(\ref{eq10}). 
For these reasons,
in Ref. \cite{Engelhardt:2015xja},
lattice QCD data have been obtained for the generalized Boer-Mulders shift
(\ref{gbmshift}), at finite $b_T$, using a pion mass
of 518 MeV.
\begin{figure}
[ptb]
\begin{center}
\includegraphics[width=13cm]{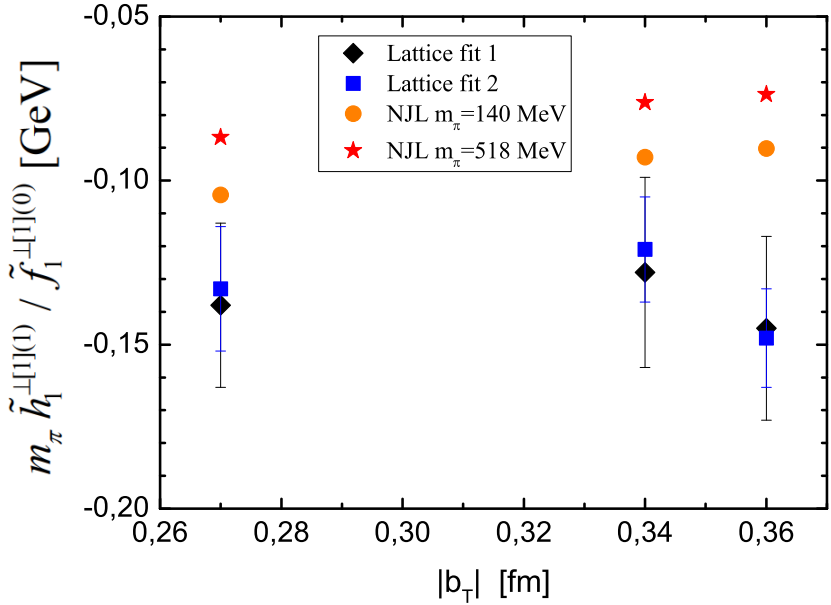}
\caption{
The generalized
Boer-Mulders shift, Eq. (\ref{gbmshift}), as a function
of $b_T$.
The orange circles are obtained using the present NJL approach with
$m_\pi=140$ MeV; the red stars are obtained using $m_\pi=518$ MeV,
the value at which the lattice measurement has been performed. 
Two different sets of lattice data (black and blue boxes), 
with their RMS deviation,
obtained in Ref. 
\cite{Engelhardt:2015xja} using independent fits, are shown for comparison.
}%
\label{f15}%
\end{center}
\end{figure}

In the following, we compare these lattice results with the outcome 
of our approach.
The numerator and denominator of the generalized
BM shift, Eq. (\ref{gbmshift}), i.e., the moments
${\tilde{f}_{1}^{[1](0)} (b_T^2 )}$ and
$m_\pi \tilde{h}_1^{\perp [1] (1)}(b_T^2 )$, 
are shown in Figs. \ref{f13} and \ref{f14},
respectively, as a function
of $b_T$, for $m_\pi=140$ MeV (full line) and $m_\pi=518$ MeV
(dashed line).
In Fig. \ref{f13} we observe that, by increasing the pion mass,
${\tilde{f}_{1}^{[1](0)} (b_T^2 )}$ is shifted towards higher values of
$b_T$. This is consistent with the fact that the e.m. radius of the pion
is smaller for $m_\pi=140$ MeV than for $m_\pi=518$ MeV
(see the Appendix for the actual values).
In Fig. \ref{f14} we compensated an overall factor $m_\pi^{-1}$ 
present in $\tilde{h}_1^{\perp [1] (1)}(b_T^2 )$
multiplying the latter quantity by the corresponding pion mass. 
We observe the same behavior, in relation with the variation of the mass, 
as in Fig. \ref{f13}. 
Nevertheless, it is difficult to give any simple intuitive explanation 
because here we are dealing with a two body operator, 
as it can be seen from Fig. \ref{f7} or Eq. (\ref{2body}).

In Ref. \cite{Engelhardt:2015xja}, lattice data
for the generalized Boer-Mulders shift have been presented
for three different 
values of $b_T$, at the momentum scale $Q =2$ GeV, given in 
\cite{Musch:2010ka}.
Our model results have therefore to be evolved to this scale,
for a proper comparison with the lattice calculation.
Unfortunately, the complete QCD evolution
of the moments
${\tilde{f}_{1}^{[1](0)} (b_T^2 )}$ and $h_1^{\perp [1] (1)}(b_T^2 )$,
involving both the $x$ and $k_T$ dependencies, is not under
control. In particular, the evolution of the $k_T$ dependence
is basically unknown.
To estimate the evolution of the $x$ dependence of the denominator,
we can approximate it with the behavior of the corresponding 
$k_T$ integrated quantity, the first moment of the PD, 
which does not evolve in $x$.
For the numerator, being $h_1^{\perp [1] (1)}(b_T^2 )$
proportional to $\alpha_s(Q^2)$ 
through the BM function,
for a first estimate
one can assume that its evolution 
is basically governed by that
of $\alpha_s(Q^2)$.
It is therefore important to fix properly the value of 
$\alpha_s$ in evaluating the model prediction
for the generalized Boer-Mulders shift at 2 GeV.
Following the discussion on the fixing of the LO evolution
parameters in NJL calculations of parton distributions,
reported in the previous section,
we use $\alpha_s$ = 0.31.


\begin{table}[tbp] \centering
\begin{tabular}
[c]{|c|c|c|c|c|}\hline
$\left\vert b_{T}\right\vert $ & Lattice 1 & Lattice 2 & NJL$~m_{\pi}=140$
MeV & NJL$~m_{\pi}=518$ MeV\\
fm & GeV & GeV & GeV & GeV\\\hline
0.27 & -0.138(28) & -0.133(19) & -0.104 & -0.087\\
0.34 & -0.128(29) & -0.121(16) & -0.093 & -0.076\\
0.36 & -0.145(25) & -0.148(15) & -0.090 & -0.074\\\hline
\end{tabular}
\caption{The generalized Boer-Mulders shift, Eq. (\ref{gbmshift}), 
as a function of $b_T$.
The first column corresponds to the three values of $b_T$ analyzed in 
Ref. \cite{Engelhardt:2015xja}.
The second and third columns contain the two different sets of lattice data, 
with their RMS deviation, obtained
in Ref. \cite{Engelhardt:2015xja} using independent fits.
In the fourth and fifth columns the results obtained using the present 
NJL approach are given, with $m_\pi=140$ MeV
and $m_\pi=518$ MeV, 
the value at which the lattice measurement has been performed.}\label{Table1}
\end{table}


In Fig. \ref{f15} and Tab. \ref{Table1} our results are compared 
with the lattice data, evaluated according
to two different fits providing consistent results [6]. We obtain a reasonably
good agreement. It must be emphasized that our
calculation has been performed in the NJL model without introducing any
new parameter. We observe that the generalized Boer-Mulders shift varies
slowly when we go from $m_{\pi}=140$ MeV to 
$m_{\pi}=518$ MeV.

The main uncertainty in our calculation comes from the poorly 
known QCD evolution of the moments of the TMDs,
entering the definition of the generalized Boer-Mulders shift.
Summarizing our approximated evolution scheme,
the $x$ evolution of the denominator has been neglected
thinking to the behavior of the corresponding PD,
the one of the numerator has been assumed to be governed by
that of $\alpha_s$ only, and the $k_T$ evolution has been
neglected overall.

\section{Conclusions}

We have considered the well-established NJL model, without any 
additional parameter, for the study of the two leading twist pion TMDs, 
the unpolarized,
$f_{1,\pi}(x,k_{T}^{2})$, and the Boer-Mulders one, 
$h_{1,\pi}^{\perp}(x,k_{T}^{2})$. We were 
motivated by the success of this model in 
reproducing pion observables, such as
the parton distribution and the pion
gamma transition form factor, and by the aim
of reproducing recent lattice results
\cite{Engelhardt:2015xja}.
Since in the latter calculation a value of 518 MeV has been used for 
the pion mass, we present our results for $m_\pi=140$ MeV and 
$m_\pi=518$ MeV.

We have studied the $k_{T}$ dependence of $f_{1,\pi}(x,k_{T}^{2})$ and
$h_{1,\pi}^{\perp}(x,k_{T}^{2})$. In both cases, this dependence is
automatically generated by the NJL dynamics. The obtained 
$k_{T}$ asymptotic behavior
of these two quantities, at the momentum scale of the model, $Q_{0}$,
is found to be $k_{T}^{-6}$ and $k_{T}^{-4}$,
respectively. Nevertheless, QCD evolution to higher scales
could modify these trends.

We observe a soft dependence on $x$ in both TMDs at $Q_{0}$.
This can be easily understood observing that, in the final expressions 
of the TMDs, the $x$-dependent part is always multiplied by $m_{\pi}$. 
Our experience with the parton distribution and the
distribution amplitude of the pion is that this $x$ dependence 
provides remarkably good results
after QCD evolution.
When  the $m_{\pi}=518$ MeV case is considered,
we get a stronger $x$ dependence, approaching results obtained in
models built with constituent quarks. 

Finally, we have studied the generalized Boer-Mulders shift,
which has been recently calculated.
The agreement we obtain with these lattice data is rather
good, qualitatively and quantitatively. 
Our results show a weak dependence on the mass of the pion.

The main theoretical uncertainty in our calculation comes from
the approximated QCD evolution we have performed.
A more conclusive comparison would require therefore 
further lattice data and the implementation of the correct evolution
of the TMDs moments appearing in the calculation.

\section{Acknowledgments}

This work was supported in part by the Mineco
under contract FPA2013-47443-C2-1-P, by GVA-Prometeo/II/2014/066, by
CPAN(CSD- 00042) and by
the Centro de Excelencia Severo Ochoa Programme grant SEV-2014-0398.
S.S. thanks the Department of Theoretical Physics of the
University of Valencia for warm hospitality and support. S.N. thanks the INFN,
sezione di Perugia, the University of Perugia 
and the Department of Physics and Geology of the University
of Perugia for warm hospitality and support.

\appendix{}

\section{The NJL model and regularization scheme}

The Lagrangian density in the two-flavor version of the NJL model with 
electromagnetic (e.m.) coupling is%
\[
\mathcal{L}=\bar{\psi}\left(  i\,\not D  -m_{0}\right)  \psi+g\left[  \left(
\bar{\psi}\,\psi\right)  ^{2}+\left(  \bar{\psi}\,\vec{\tau}\,i\gamma
_{5}\,\psi\right)  ^{2}\right]
\]
with $D_{\mu}=\partial_{\mu}+i\,e\,A_{\mu}.$ 

The NJL model is a
non-renormalizable field theory and a cut-off procedure must be defined. We
have used the Pauli-Villars regularization in order to render the occurring
integrals finite. This means that, for any integral, we make the replacement%
\[
\int\frac{d^{4}q}{\left(  2\pi\right)  ^{4}}f\left(  q;m\right)
~\longrightarrow~\int\frac{d^{4}q}{\left(  2\pi\right)  ^{4}}\,\sum_{j=0}%
^{2}\,c_{j}\,f\left(  q;M_{j}\right)
\]
with $M_{j}^{2}=m^{2}+j\,\Lambda^{2}$, $c_{0}=c_{2}=1$ and $c_{1}=-2$.
Following ref. \cite{Klevansky:1992qe} we determine the regularization
parameter $\Lambda$ and $m$ by calculating the pion decay constant and the
quark condensate in the chiral limit, via%
\[
f_{\pi}^{2}=-\frac{3\,m^{2}}{4\,\pi^{2}}\sum_{j=0}^{2}\,c_{j}\,\log\frac
{M_{j}^{2}}{m^{2}}~,~~~\ \ \ \ \ \ \ \ ~\left\langle \bar{u}u\right\rangle
=-\frac{3\,m}{4\,\pi^{2}}\sum_{j=0}^{2}\,c_{j}\,M_{j}^{2}\,\log\frac{M_{j}%
^{2}}{m^{2}}%
\]
with $m_{0}$ fixing the pion mass.

With the conventional values $\left\langle \bar{u}u\right\rangle =-(0.250%
\operatorname{GeV}%
)^{3},$ $f_{\pi}=0.0924%
\operatorname{GeV}%
$ and $m_{\pi}=0.140%
\operatorname{GeV}%
$, we get $m=0.238%
\operatorname{GeV}%
$, $\Lambda=0.860%
\operatorname{GeV}%
$ and $m_{0}=5.4%
\operatorname{MeV}%
.$ For the pion-quarks coupling constant we get $g_{\pi qq}^{2}=6.279$.
The electromagnetic  
pion radius turns out to be $r_{\pi}^{2}=0.31%
\operatorname{fm}%
^{2}$ (experimental value $r_{\pi}^{2}=0.44%
\operatorname{fm}%
^{2}$).

For a proper comparison  with lattice data, we have applied the same model to
a massive pion, with $m_{\pi}=0.518%
\operatorname{GeV}%
.$ We have not changed the value of $\Lambda$; 
for $m$ we have taken $m=0.300%
\operatorname{GeV}%
.$ In this way we have $\left\langle \bar{u}u\right\rangle =-(0.263%
\operatorname{GeV}%
)^{3},$ $f_{\pi}=0.0997%
\operatorname{GeV}$,
$m_0 = 83$ MeV and $g_{\pi qq}^{2}=3.667.$ 
The e.m. pion radius is $r_{\pi}^{2}=0.38%
\operatorname{fm}%
^{2}.$

\end{document}